# An Industrial Case Study on Test Cases as Requirements

Elizabeth Bjarnason[1], Michael Unterkalmsteiner[2], Emelie Engström[1], Markus Borg[1]

[1]Lund University
SE-221 00 Lund, Sweden
FirstName.LastName@cs.lth.se

[2]Blekinge Institute of Technology
SE-371 79 Karlskrona, Sweden
mun@bth.se

**Abstract.** It is a conundrum that agile projects can succeed 'without requirements' when weak requirements engineering is a known cause for project failures. While Agile development projects often manage well without extensive requirements documentation, test cases are commonly used as requirements. We have investigated this agile practice at three companies in order to understand how test cases can fill the role of requirements. We performed a case study based on twelve interviews performed in a previous study. The findings include a range of benefits and challenges in using test cases for eliciting, validating, verifying, tracing and managing requirements. In addition, we identified three scenarios for applying the practice, namely as a mature practice, as a de facto practice and as part of an agile transition. The findings provide insights into how the role of requirements may be met in agile development including challenges to consider.

## 1 Introduction

Agile development methods strive to be responsive to changing business requirements by integrating requirements, design, implementation and testing processes [1][2]. Face-to-face communication is prioritised over written requirements documentation and customers are expected to convey their needs directly to the developers [3][4]. However, weak customer communication in combination with minimal documentation is reported to cause problems in scaling and evolving software for agile projects [4].

Requirements specifications fill many roles. They are used to communicate among stakeholders within a software development project, to drive design and testing, and to serve as a reference for project managers and in the evolution of the system [6]. Due to the central role of requirements in coordinating software development, there exists a plethora of research on how to document requirements with varying degrees of formality depending on its intended use. This spans from formal requirements specifications [7] and requirements models [8], over templates [9] to user stories [10] and requirements expressed using natural language. At the formal end of the spectrum, requirements specifications can be automatically checked for consistency [11] and used to derive other artefacts, e.g. software designs [12] or test cases [13]. For the less formal approaches, requirements documentation is driven by heuristics and best practices for achieving high quality [14] requirements.

The coordination of evolving requirements poses a challenge in aligning these with later development activities including testing [5]. In a previous study we identified the use of test cases as requirements (TCR) as one of several industrial practices used to address this challenge [5]. In this paper, we investigate this practice further by a more detailed analysis of the interview data from the three case companies (of six) that explicitly mentioned this practice. The case study presented in this paper investigates how the practice may support the role of requirements engineering (RE) by investigating **RQ1** How does the TCR practice fulfil the role of requirements? and **RQ2** Why and how is the TCR practice applied?

The rest of this paper is organized as follows. Section 2 describes related work. Section 3 presents the case companies and Section 4 the applied research method. The results are reported in Section 5, while the research questions are answered in Sections 6 and 7. The paper is concluded in Section 8.

## 2 Agile RE: Test Cases as Requirements Documentation

In agile software development requirements and tests can be seen as two sides of the same coin. Martin and Melnik [15] hypothesize that as the formality of specifications increases, requirements and tests become indistinguishable. This principle is taken to the extreme by unit tests [16] where requirements are formalized in executable code. Practitioners report using unit tests as a technical specification that evolves with the implementation [17]. However, unit tests may be too technical for customers and thereby lack the important attribute of being understandable to all relevant stakeholders.

Acceptance tests are used to show customers that the system fulfils the requirements [18]. However, developing acceptance tests from requirements specifications is a subjective process that does not guarantee that all requirements are covered [18]. This is further complicated by requirements documentation rarely being updated [19], leading to potentially outdated acceptance tests. In agile development, automated acceptance tests (AATs) drive the implementation and address these issues by documenting requirements and expected outcomes in an executable format [4][20]. This agile practice is known, among others, as customer tests, scenario tests, executable/automated acceptance tests, behaviour driven development and story test driven development [21].

Some organisations view and use the AATs as requirements thereby fully integrating these two artefacts [15]. AATs are used to determine if the system is acceptable from a customer perspective and used as the basis for customer discussions, thus reducing the risk of building the wrong system. However, the communication might be more technical and require more technical insight of the customer. Melnik et al. [22] found that customers in partnership with software engineers could communicate and validate business requirements through AATs, although there is an initial learning curve.

The conceptual difficulty of specifying tests before implementation [23][24][25] led to the conception of behaviour-driven development (BDD) [26]. BDD incorporates aspects of requirements analysis, requirements documentation and communication, and automated acceptance testing. The behaviour of a system is defined in a domain-specific language (DSL); a common language that reduces ambiguities and misunderstandings. This is further enhanced by including terms from the business domain in the DSL.

Haugset and Hansen studied acceptance test driven development (ATDD) as an RE practice and report on its benefits and risks [20]. Our work extends on this by also investigating companies that use the TCR practice without applying ATDD principles.

## 3   Case Companies

The three case companies all develop software using an agile development model. However, a number of other factors vary between the companies. These factors are summarised in **Table 1** and the interviewees are characterised in **Table 2.** .

Table 1. Overview of the case companies.

| Company | A | B | C |
|---|---|---|---|
| **Type of company** | Softw. develop., embedded products | Consulting | Softw. develop., embedded products |
| **#employees in softw development** | 125-150 | 135 | 1,000 |
| **#employees in typical project** | 10 | Mostly 4-10, but varies greatly | Previously: 800-1,000 person years |
| **Distributed** | No | No | Yes |
| **Domain / System type** | Computer networking equipment | Advisory/technical services, appl. management | Telecom |
| **Source of reqts** | Market driven | Bespoke | Bespoke, market driven |
| **Main quality focus** | Availability, performance, security | Depends on customer focus | Performance, stability |
| **Certification** | Not software related | No | ISO9001 |
| **Process Model** | Agile | Agile in variants | Agile with gate decisions Previous: Waterfall |
| **Project duration** | 6-18 months | No typical project | Previously: 2 years |
| **#requirements in typical project** | 100 (20-30 pages HTML) | No typical project | Previously: 14,000 |
| **#test cases in typical project** | ~1,000 test cases | No typical project | Previously: 200,000 for platform, 7,000 for system |
| **Product Lines** | Yes | No | Yes |
| **Open Source** | Yes | Yes incl. contributions | Yes (w agile dev model) |

### 3.1   Company A

Company A develops network equipment consisting of hardware and software. The software development unit covered by the interview study has around 150 employees. The company is relatively young but has been growing fast during the past few years. A typical software project has a lead time of 6-18 months, around 10 co-located members and approximately 100 requirements and 1,000 system test cases. A market-driven requirements engineering process is applied. The quality focus for the software is on availability, performance and security. Furthermore, the company applies a product-line approach and uses open-source software in their development.

A product manager, a project manager, and a tester were interviewed at Company A, all of which described how the company manages requirements as test cases.

### 3.2 Company B

Company B is a consultancy firm that provides technical services to projects that vary in size and duration. Most projects consist of one development team of 4-10 people located at the customer site. The requirements are defined by a customer (bespoke).

The three consultants that were interviewed at Company B can mainly be characterised as software developers. However, they all typically take on a multitude of roles within a project and are involved throughout the entire lifecycle. All three of these interviewees described the use of the TCR practice.

### 3.3 Company C

Company C develops software for embedded products in the telecommunications domain. The software development unit investigated in this study, consists of 1,000 people. At the time of the interviews, the company was transitioning from a waterfall process to an agile process. Projects typically run over 2 years and include 400-500 people. The project size and lead time is expected to decrease with the agile process. The projects handle a combination of bespoke and market-driven requirements. Including the product-line requirements, they handle a very complex and large set of requirements.

Six of the interviewees (of 15) discussed the practice, namely one requirements engineer, two project managers, two process managers and one tester.

**Table 2.** Interviewees per company. Experience in role noted as S(enior) = more than 3 years, or J(unior) = up to 3 years. Interviewees mentioning the TCR practice are marked with **bold**. Note: For Company B, software developers also perform RE and testing tasks.

| Role | A | B | C |
|---|---|---|---|
| Requirements engineer | | | **C1:S**, C6:S, C7:S |
| Systems architect | | | C4:S |
| Software developer | | **B1:J, B2:S, B3:S** | C13:S |
| Test engineer | **A2:S** | | C9:S, **C10:S**, C11:J, C12:S, C14:S |
| Project manager | **A1:J** | | **C3:J, C8:S** |
| Product manager | **A3:S** | | |
| Process manager | | | **C2:J, C5:S**, C15:J |

## 4 Method

We used a flexible exploratory case study design and process [27] consisting of four stages: 1) *Definition*, 2) *Evidence selection*, 3) *Data analysis* and 4) *Reporting*.

**Definition of Research Questions and Planning.** Since we were interested in how agile development can be successful 'without requirements' we selected to focus on the practice of using test cases as requirements. We formulated the research questions, (RQ1) How does the TCR practice fulfil the role of requirements? and (RQ2) Why and how is the TCR practice applied?

**Evidence Selection.** We selected to use word-by-word transcriptions from our previous study of RE-Testing coordination. The research questions of this paper are within the broader scope of the previous study [5], which also included agile processes. In addition, the semi-structured interviews provided rich material since the interviewees could freely describe how practices were applied including benefits and challenges. Data selection was facilitated by the rigorous coding performed in the previous study. We selected the interview parts coded for the TCR practice. In addition, the transcripts were searched for key terms such as 'acceptance test', 'specification'.

**Data Analysis.** The analysis of the selected interview data was performed in two steps. First the transcripts were descriptively coded. These codes were then categorised into benefits and challenges, and reported per case company in Section 5. The analysis was performed by the first author. The results were validated independently by the third author. The third author analysed and interpreted a fine-grained grouping of the interview data produced in the previous study, and compared this against the results obtained by the first researcher. No conflicting differences were found.

## 5 Results

Two of the investigated companies apply the TCR practice while the third company plan to apply it. The maturity of the practice thus varied. The interviewees for Company B provided the most in depth description of the practice, which is reflected in the amount of results per company. Limitations of the findings are discussed in Section 5.4.

### 5.1 Company A: A De Facto Practice

Test cases have become the de facto requirements in company A due to weak RE (A2[1]), i.e. the RE maturity in the company is low while there is a strong competence within testing. Formal (traditional) requirements are mainly used at the start of a project. However, these requirements are not updated during the project and lack traceability to the test cases. Instead, the test cases become the requirements in the sense that they verify and ensure that the product fulfils the required behaviour.

**Benefits.** Efficient way of managing requirements in a small and co-located organisation that does not require managing and maintaining a formal requirements specification once test design has been initiated (A1). In addition, the structure of the test specifications is closer to the code simplifying navigation of these 'requirements' once the implementation has started (A1).

---

[1] Mentioned by this interviewee, see interviewee codes in Table 2.

**Challenges.** As the company grows, the lack of traces to formal requirements is a problem in communication of requirements changes to the technical roles (A1, A2) and in ensuring correct test cases (A2). In addition, the test cases lack information about requirements priority, stakeholders etc., needed by the development engineers when a test case fails (A2) or is updated (A3). The untraced artefacts do not support either ensuring test coverage of the formal requirements (A1, A3), or identifying the test cases corresponding to the requirements re-used for a new project (A2).

### 5.2 Company B: An Established Practice

Company B actively applies the TCR practice through behaviour-driven development supported by tools. The customer and the product owner define product and customer requirements. Then, for each iteration, the development engineers produce acceptance criteria (user scenarios) and acceptance test cases from these requirements. These 'requirements test cases' are iterated with the business roles to ensure validity (B1), and entered into an acceptance test tool that produces an executable specification. The interviewees described that the acceptance criteria can be used as a system specification. However, interviewee B3 stated that the acceptance criteria can be read 'to get an impression. But, if you wonder what it means, you can look at the implementation', i.e. this documentation is not fully stand-alone.

**Benefits.** The interviewees stated that the main benefits are improved customer collaboration around requirements, strengthened alignment of business requirements with verification, and support for efficient regression testing. The customer collaboration raises the technical discussion to a more conceptual level while also improving requirements validity, since, as an engineer said, 'we understand more of the requirements. They concretize what we will do.' (B1) This alignment of business and technical aspects was experienced to also be supported when managing requirements changes by the use of acceptance test cases as formal requirements (B2, B3). At the end of a project the acceptance test cases show 'what we've done' (B2). Furthermore, the executable specification provided by this practice, in combination with unit tests, acts as a safety net that enables projects to 'rebound from anything' (B1) by facilitating tracking of test coverage, efficiently managing bugs and performance issues.

**Challenges.** The interviewees mentioned several challenges for the practice concerning active customer involvement, managing complex requirements, balancing acceptance vs. unit tests and maintaining the 'requirements test cases'. Over time the company has achieved active customer involvement in defining and managing requirements with this practice, but it has been challenging to ensure that 'we spoke the same language' (B3). The interviewees see that customer competence affects the communication and the outcome. For example, interviewee B3 said that non-technical customers seldom focus on quality requirements. Similarly, getting the customer to work directly with requirements (i.e. the acceptance test cases) in the tool has not been achieved. This is further complicated by issues with setting up common access across networks.

Complex interactions and dependencies between requirements, e.g. for user interfaces (B1) and quality requirements (B2), are a challenge both to capture with ac-ceptance test cases and in involving the customer in detailing them. Furthermore, automatically testing performance and other quality aspects on actual hardware and in a live testing environment is challenging to manage with this approach.

All interviewees mentioned the challenge in balancing acceptance vs. unit test cases. It can be hard to motivate engineers to write acceptance-level test cases. Furthermore, maintenance of the acceptance test cases needs to be considered when applying this practice (B1, B2, B3). Interviewee B3 pointed out that test cases are more rigid than requirements and thus more sensitive to change. There is also a risk of deteriorating test case quality when testers make frequent fixes to get the tests to pass (B2).

### 5.3 Company C: Planned Practice as part of Agile Transition

The agile transition at the company included introduction of this practice. Requirements will be defined by a team consisting of a product owner, developers and testers. User stories will be detailed into requirements that specify 'how the code should work' (C8). These will be documented as acceptance test cases by the testers and traced to the user stories. Another team will be responsible for maintaining the software including the user stories, test cases and traces between them. In the company's traditional process, test cases have been used as quality requirements, as a de facto practice. Interviewee C1 describes an attempt to specify these as formal requirements that failed due to not reaching an agreement on responsibility for the cross-functional requirements within the development organisation.

**Benefits.** The practice is believed to decrease misunderstandings of requirements between business and technical roles, improve on the communication of changes and in keeping the requirements documentation updated (C5, C10).

**Challenges.** Integrating the differing characteristics and competences of the RE and testing activities are seen as a major challenge (C5, C10) in the collaboration between roles and in the tools. RE aspects that need to be provided in the testing tools include noting the source of a requirement, connections and dependencies to other requirements and validity for different products (C5).

### 5.4 Limitations

We discuss limitations of our results using guidelines provided by Runeson et al. [27].

**Construct validity.** A main threat to validity lies in that the analysed data stems from interviews exploring the broader area of coordinating RE and testing. This limits the depth and extent of the findings to what the interviewees spontaneously shared around the practice in focus in this paper. In particular, the fact that the practice was not yet fully implemented at Company C at the time of the interviews limits the insights gained from those interviews. However, we believe that the broad approach of the original study in combination with the semi-structured interviews provide valuable insights, even though further studies are needed to fully explore the topic.

**External validity.** The findings may be generalized to companies with similar characteristics as the case companies (see Section 3), by theoretical generalization [27].

**Reliability.** The varying set of roles from each case poses a risk of missing important perspectives, e.g. for Company B the product owner's view would complement the available interview data from the development team. There is a risk of researcher bias in the analysis and interpretation of the data. This was partly mitigated by triangulation; two researchers independently performing these steps. Furthermore, a rigorous process was applied in the (original) data collection including researcher triangulation of interviewing, transcription and coding, which increases the reliability of the selected data.

## 6 Test Cases in the Role of Requirements (RQ1)

We discuss how the TCR practice supports the main roles of RE and the requirements specification according to roles defined by Lauesen [28], i.e. the elicitation and validation of stakeholders' requirements; software verification; tracing; and managing requirements. The discussion is summarised in **Table 3**.

**Table 3.** Summary of benefits and challenges per role of RE.

| Benefits | Challenges |
|---|---|
| Elicitation and Validation | |
| Cross-functional communication | Good Customer-Developer relationship |
| Align goals & perspectives between roles | Active customer involvement |
| Address barrier of specifying solutions | Sufficient technical and RE competence |
|  | Complex requirements |
| Verification | |
| Supports regression testing | Quality requirements |
| Increased requirements quality |  |
| Test coverage |  |
| Tracing | |
| Requirements - test case tracing in BDD | Tool integration |
| Requirements Management | |
| Maintaining RET alignment | Locating impacted requirements |
| Requirement are kept updated | Missing requirement context |
| Communication of changes | Test case maintenance |
| Efficient documentation updates |  |

### 6.1 Elicitation and Validation

The TCR practice supports elicitation and validation of requirements by its direct and frequent communication between business and technical roles for all companies. The customer involvement in combination with increased awareness of customer perspectives among the technical roles supports defining valid requirements. This confirms observations by Melnik and Maurer [29], Park and Maurer [30], Haugset and Hanssen [20] and Latorre [31]. Furthermore, at Company B, the use of the acceptance criteria format led to customers expressing requirements at a higher abstraction level instead of focusing on technical details. Thus, this practice can address the elicitation barrier of requesting specific solutions rather than expressing needs [28].

Nevertheless, the TCR practice requires good customer relations, as stated by interviewees in Company B. Active customer involvement is a known challenge for agile RE due to time and space restrictions for the customer, but also due to that this role requires a combination of business and technical skills [4][31]. Business domain tools can be used to facilitate the customers in specifying acceptance tests [30]. For example, Haugset and Hanssen [20] report that customers used spread-sheets to communicate information and never interacted directly with actual test cases.

Eliciting and validating requirements, in particular complex ones, relies on competence of the roles involved. At Company B limited technical knowledge affected the customer's ability to discuss quality requirements. This can lead to neglecting to elicit them altogether [4]. Similarly, capturing complex requirements with acceptance test cases is a challenge, in particular for user interactions and quality requirements.

### 6.2 Verification

The TCR practice supports verification of requirements by automating regression tests as for Company B. The AATs act as a safety net that catches problems and enables frequent release of product-quality code. This was also observed by Kongsli [32], Haugset and Hanssen [20], and Latorre [31]. The practice ensures that all specified requirements (as test cases) are verified and test coverage can be measured by executing the tests.

The verification effort relies on verifiable, clear and unambiguous requirements [6]. Test cases are per definition verifiable and the format used by Company B supports defining clear requirements. Nevertheless, Company B mentioned quality requirements as a particular challenge for embedded devices as this requires actual hardware. This confirms previous findings by Ramesh [4] and Haugset and Hanssen [20] that quality requirements are difficult to capture with AATs.

### 6.3 Tracing

Tracing of requirements and test cases is supported by the TCR practice, however the benefits depend on the context. Merely using test cases as de facto requirements (as in Company A) does not affect tracing. For the BDD approach applied at Company B, the tools implicitly trace acceptance criteria and test cases, although there are no traces between the original customer requirements and the acceptance criteria. Hence, as the requirements evolve [33] this knowledge is reflected purely in the test cases.

At Company C, where user stories were to be detailed directly into acceptance test cases, tracing remains a manual, albeit straight forward task of connecting acceptance test cases to the corresponding user stories. Furthermore, the responsibility for these traces is clearly defined in the development process, a practice identified by Uusitalo [34] as supporting traceability. However, it is a challenge for the company to identify tools which provide sufficient support for requirements and for testing aspects, and for the integration of the two.

### 6.4 Requirements Management

The TCR practice provides benefits in managing requirements in an efficient way throughout the life-cycle. As mentioned for Companies A and B, the practice facilitates a joint understanding of requirements that provides a base for discussing and making decisions regarding changes. However, the practice also requires effort in involving development engineers in the requirements discussion. The optimal balance between involving these technical roles to ensure coordination of requirements versus focusing on pure development activities remains as future work.

The challenge of keeping requirements updated after changes [5] is addressed by a close integration with test cases, as for Company B, since the test cases are by necessity updated throughout the project. Furthermore, since the requirements are documented in an executable format, conflicting new or changed requirements are likely to cause existing test cases to fail. However, locating requirements in a set of test cases was mentioned as a challenge for Company B due to badly structured test cases. The difficulty of organizing and sorting automated tests has also been reported by Park [21].

Contextual requirements information, e.g. purpose and priority [28], is seldom retained in the test cases but can support, for example, impact analysis and managing failed test cases. Without access to contextual information from the test cases, additional effort is required to locate it to enable decision making.

## 7 The Reasons for and Contexts of the Practice (RQ2)

Each case company applies the practice differently and for different reasons. At Company A it has become a *de facto practice* due to strong development and test competence, and weak RE processes. However, merely viewing test cases as requirements does not fully compensate for a lack of RE. Company A faces challenges in managing requirements changes and ensuring test coverage of requirements. The requirements documentation does not satisfy the information needs of all stakeholders and staff turnover may result in loss of (undocumented) product knowledge. As size and complexity increase so does the challenge of coordinating customer needs with testing effort [5].

Company B applies the practice *consciously using a full BDD approach including tool support*. This facilitates customer communication through which the engineering roles gain requirements insight. The AATs provide a feedback system confirming the engineers' understanding of the business domain [30]. However, it is a challenge to get customers to specify requirements in the AAT tools. Letting domain experts or customers provide information via e.g. spread-sheets may facilitate this [30].

The third practice variant is found at Company C, where it is consciously *planned as part of a transition to agile processes* applying story test driven development [21]. The practice includes close and continuous collaboration around requirements between business and development roles. However, no specific language for expressing the acceptance criteria or specific tools for managing these are planned. In contrast to the de facto context, Company C envisions this practice as enabling analysis and maintenance of requirements. To achieve this, requirements dependencies and priorities need to be supported by the test management tools.

## 8 Conclusions and Future Work

Coordinating and aligning frequently changing business needs is a challenge in software development projects. In agile projects this is mainly addressed through frequent and direct communication between the customer and the development team, and the detailed requirements are often documented as test cases.

Our case study provides insights into how this practice meets the various roles that the requirements play. The results show that the direct and frequent communication of this practice supports eliciting, validating and managing new and changing customer requirements. Furthermore, specifying requirements as acceptance test cases allow the requirements to become a living document that supports verifying and tracing requirements through the life cycle. We have also identified three contexts for this practice; as a de facto practice, part of an agile transition and as a mature practice.

The results can aid practitioners in improving their agile practices and provide a basis for further research. Future work includes investigating how to further improve the RE aspects when documenting requirements as test cases.

**Acknowledgement.** We want to thank the interviewees. This work was funded by EASE (ease.cs.lth.se).